\documentclass[a4paper,11pt,fleqn]{article}
\usepackage{amsmath,cite}
\usepackage[hidelinks]{hyperref}

\begin{document}

\title{\textbf{ On continual classes of evolution equations}}

\author{\textsc{Sergei Sakovich}\bigskip \\
\small Institute of Physics, National Academy of Sciences of Belarus \\
\small \href{mailto:sergsako@gmail.com}{sergsako@gmail.com}}

\date{}

\maketitle

\begin{abstract}
The original Miura transformation, considered as a nonlinear potential transformation, is applicable to a continual class of evolution equations, not only to discrete integrable equations and their hierarchies. The same continual class of evolution equations appears from a different problem, namely, from the gauge-invariant description of a zero-curvature representation with a definite $x$-part containing no essential parameter.
\end{abstract}

\section{Introduction}

The original Miura transformation was introduced in \cite{M68} as a nonlinear potential transformation relating the Korteweg--de~Vries (KdV) equation to the modified KdV equation. The Miura transformation is applicable not only to the KdV equation and its integrable hierarchy. For example, the Sawada--Kotera equation and the Kaup--Kupershmidt equation also admit the Miura transformation \cite{FG80}. Moreover, as we have shown in \cite{S88}, the Miura transformation is applicable to a very wide (continual) class of evolution equations, and most of those equations, of course, should be non-integrable in any reasonable sense. Let us note that this result of \cite{S88} was inspired by the Lie--B\"{a}cklund algebra structure of the exactly solvable Liouville equation.

In the present paper, in Section~\ref{s2}, we recall the direct derivation of the complete class of evolution equations admitting the Miura transformation, without any reference to Lie--B\"{a}cklund algebras. In Section~\ref{s3}, we show that exactly the same continual class of evolution equations appears from a different problem, namely, from the gauge-invariant description of a zero-curvature representation with a definite $x$-part containing no essential parameter. In Section~\ref{s4}, we comment on the obtained results.

\section{The Miura transformation} \label{s2}

Let us find all the local evolution equations
\begin{equation}
u_t = f ( x , t , u , u_1 , \dotsc , u_n ) \label{e1}
\end{equation}
which admit the Miura transformation
\begin{equation}
u = v_1 - \frac{1}{2} v^2 , \label{e2}
\end{equation}
where $u_i$ and $v_i$ ($i=1,2,\dotsc$) denote $\partial_x^{\,i} u$ and $\partial_x^{\,i} v$, respectively.

The Miura transformation \eqref{e2} relates a local evolution equation \eqref{e1} to a local evolution equation
\begin{equation}
v_t = g ( x , t , v , v_1 , \dotsc , v_n ) \label{e3}
\end{equation}
if the function $u(x,t)$ determined by \eqref{e2} satisfies \eqref{e1} whenever the function $v(x,t)$ satisfies \eqref{e3}. This can be equivalently expressed by the condition
\begin{equation}
f ( x , t , u , u_1 , \dotsc , u_n ) = \bigl( D_x - v \bigr) g ( x , t , v , v_1 , \dotsc , v_n ) , \label{e4}
\end{equation}
where $D_x$ denotes the total derivative with respect to $x$. Note that the condition \eqref{e4} must be a differential consequence of the relation \eqref{e2}, otherwise \eqref{e4} and \eqref{e2} would produce an ordinary differential equation restricting solutions $v$ of the evolution equation \eqref{e3}.

Now, using the relation \eqref{e2} and its differential consequences, namely, $v_1 = u + \tfrac{1}{2} v^2$, $v_2 = u_1 + u v + \frac{1}{2} v^3$, $v_3 = u_2 + u^2 + u_1 v + 2 u v^2 + \frac{3}{4} v^4$, etc., we can eliminate $v_1 , v_2, \dotsc , v_{n+1}$ from the condition \eqref{e4}, and in this way we rewrite \eqref{e4} in the following equivalent form:
\begin{equation}
f ( x , t , u , u_1 , \dotsc , u_n ) = \bigl( \widetilde{D}_x - v \bigr) h ( x , t , v , u, u_1 , \dotsc , u_{n-1} ) , \label{e5}
\end{equation}
where
\begin{equation}
\widetilde{D}_x = \partial_x + \bigl( u + \tfrac{1}{2} v^2 \bigr) \partial_v + u_1 \partial_u + u_2 \partial_{u_1} + u_3 \partial_{u_2} + \dotsb \label{e6}
\end{equation}
and
\begin{equation}
h ( x , t , v , u, u_1 , \dotsc , u_{n-1} ) = g \bigl( x , t , v , \widetilde{D}_x v , \widetilde{D}_x^2 v , \dotsc , \widetilde{D}_x^n v \bigr) . \label{e7}
\end{equation}
The crucial point is that the condition \eqref{e5} must be an identity, because it cannot be a differential consequence of the relation \eqref{e2}. Therefore, the right-hand side of \eqref{e5} must be independent of $v$, this determines admissible functions $h$, and then \eqref{e5} is simply a definition of admissible functions $f$.

Next, repeatedly applying $\partial_v$ to \eqref{e5} three times and using the evident identity
\begin{equation}
\partial_v \widetilde{D}_x = \bigl( \widetilde{D}_x + v \bigr) \partial_v , \label{e8}
\end{equation}
we obtain the auxiliary condition
\begin{equation}
\bigl( \widetilde{D}_x + 2 v \bigr) \partial_v^{\,3} h = 0 . \label{e9}
\end{equation}
Since $h$ must be a local expression, it follows from \eqref{e9} that $\partial_v^{\,3} h = 0$, that is, the function $h$ is necessarily of the form
\begin{equation}
h = \frac{1}{2} v^2 p + v q + r , \label{e10}
\end{equation}
where $p$, $q$ and $r$ are functions of $x , t , u, u_1 , \dotsc , u_{n-1}$.

Finally, substituting the expression \eqref{e10} into the condition  \eqref{e5} and taking into account that $p$, $q$, $r$ and $f$ do not depend on $v$, we obtain the following expressions:
\begin{equation}
q = D_x p , \qquad r = \bigl( D_x^2 + u \bigr) p , \qquad f = \bigl( D_x^3 + 2 u D_x + u_1 \bigr) p , \label{e11}
\end{equation}
where $p$ is an arbitrary function of $x , t , u, u_1 , \dotsc , u_{n-3}$. The relations \eqref{e7}, \eqref{e10} and \eqref{e11} solve our problem. We have found that the local evolution equations \eqref{e1} admitting the Miura transformation \eqref{e2} constitute the continual class
\begin{equation}
u_t = \bigl( D_x^3 + 2 u D_x + u_1 \bigr) p ( x , t , u , u_1 , \dotsc , u_{n-3} ) , \label{e12}
\end{equation}
the corresponding local evolution equation \eqref{e3} being
\begin{multline}
v_t = \bigl( D_x^2 + v D_x + v_1 \bigr) p \bigl( x , t , v_1 - \tfrac{1}{2} v^2 , D_x ( v_1 - \tfrac{1}{2} v^2 ) , \dotsc , \bigr. \\
\bigl. D_x^{n-3} ( v_1 - \tfrac{1}{2} v^2 ) \bigr) , \label{e13}
\end{multline}
where the function $p$ and the order $n$ are arbitrary.

\section{The zero-curvature representation} \label{s3}

The continual class of evolution equations \eqref{e12} appears from a completely different problem as well. Let us find all the local evolution equations \eqref{e1} which admit zero-curvature representations (ZCRs)
\begin{equation}
D_t X - D_x T + [ X , T ] = 0 \label{e14}
\end{equation}
with the fixed matrix $X$ given by
\begin{equation}
X =
\begin{pmatrix}
0 & u \\
- \frac{1}{2} & 0
\end{pmatrix}
\label{e15}
\end{equation}
and any $2 \times 2$ traceless matrices $T ( x , t , u , u_1 , \dotsc , u_{n-1} )$, where $D_t$ and $D_x$ stand for the total derivatives, the square brackets denote the matrix commutator, and $u_i = \partial_x^{\,i} u$ ($i = 1, 2, \dotsc $). We solve this problem by the cyclic basis method \cite{S95,S04,KKS04,S14,S20,S23}.

In the present case of the matrix $X$ given by \eqref{e15}, the characteristic form of the ZCR \eqref{e14} of an evolution equation \eqref{e1} is
\begin{equation}
f C = \nabla T , \label{e16}
\end{equation}
where $f$ is the right-hand side of \eqref{e1}, $C = \partial X / \partial u$, and the operator $\nabla$ is defined as $\nabla M= D_x M - [ X , M ]$ for any $2 \times 2$ matrix $M$. Computing $C$, $\nabla C$, $\nabla^2 C$ and $\nabla^3 C$, we find that the cyclic basis is $\{ C, \nabla C , \nabla^2 C \}$, with the closure equation
\begin{equation}
\nabla^3 C = - u_1 C - 2 u \nabla C . \label{e17}
\end{equation}
Decomposing $T$ over the cyclic basis as
\begin{equation}
T = a_0 C + a_1 \nabla C + a_2 \nabla^2 C , \label{e18}
\end{equation}
where $a_0$, $a_1$ and $a_2$ are functions of $x , t , u , u_1 , \dotsc , u_{n-1}$, we find from \eqref{e16} and \eqref{e17} that
\begin{equation}
a_1 = - D_x a_2 , \qquad a_0 = - D_x a_1 + 2 u a_2 , \qquad f = D_x a_0 - u_1 a_2 . \label{e19}
\end{equation}
Note that the function $a_2 = p ( x , t , u , u_1 , \dotsc , u_{n-3} )$ and the order $n$ remain undetermined. The explicit expressions
\begin{equation}
T =
\begin{pmatrix}
\frac{1}{2} D_x p & D_x^2 p + u p \\[4pt]
- \frac{1}{2} p & - \frac{1}{2} D_x p
\end{pmatrix}
\label{e20}
\end{equation}
and
\begin{equation}
f = D_x^3 p + 2 u D_x p + u_1 p , \label{e21}
\end{equation}
which follow from \eqref{e18} and \eqref{e19}, solve our problem.

We have found that the local evolution equations \eqref{e1} admitting ZCRs \eqref{e14} with the matrix $X$ given by \eqref{e15} constitute the continual class \eqref{e12}, the corresponding matrices $T$ being determined by \eqref{e20}.

\section{Conclusion} \label{s4}

We have shown that the original Miura transformation \eqref{e2}, considered as a nonlinear potential transformation, is applicable to a continual class of evolution equations \eqref{e12}, not only to discrete integrable equations and their hierarchies. We have also shown that exactly the same continual class of evolution equations \eqref{e1} with \eqref{e21} appears from a different problem, namely, from the gauge-invariant description of a zero-curvature representation \eqref{e14} with the given $x$-part \eqref{e15} containing no essential parameter.

Not every expression of the form $u = s ( x , t , v , v_1 , \dotsc , v_m )$ can serve as a Miura-type transformation between two local evolution equations, as was shown by a simple polynomial generalization of the original (quadratic) Miura transformation \cite{S90}. In many interesting special cases, the general Miura-type transformations can be analyzed and represented as chains of simpler transformations \cite{S93}, and this definitely deserves further investigation. We believe that the relation between the differential substitutions (another name of Miura-type transformations) and the so-called pseudosymmetries \cite{Sok88} may be very useful, for the following reason. Pseudosymmetries correspond to ZCRs with some fixed $x$-parts \cite{Sok88}, whereas ZCRs with fixed $x$-parts always represent some continual classes of evolution equations \cite{S95}.

\end{document}